\documentclass[aps,pra,showpacs,superscriptaddress,twocolumn,tightenlines]{revtex4}
\usepackage{amsmath}
\usepackage{amsfonts}
\usepackage{amssymb}
\usepackage{graphicx}
\usepackage{epsfig}
\usepackage{color}
\usepackage[colorlinks,citecolor=blue]{hyperref}

\begin{document}

\title{A superconducting circuit probe for analog quantum simulators}

\author{Liang-Hui Du}
\affiliation{School of Natural Sciences, University of California, Merced, California 95343, USA}
\author{J. Q. You}
\affiliation{Beijing Computational Science Research Center, Beijing 100084, China}
\author{Lin Tian}
\email{ltian@ucmerced.edu}
\affiliation{School of Natural Sciences, University of California, Merced, California 95343, USA}

\begin{abstract}
Analog quantum simulators can be used to study quantum correlation in novel many-body systems by emulating the Hamiltonian of these systems. One essential question in quantum simulation is to probe the properties of an emulated many-body system. Here we present a circuit QED scheme for probing such properties by measuring the spectrum of a superconducting resonator coupled to a quantum simulator. We first study a general framework of this approach, and show that the spectrum of the resonator is directly related to the correlation function of the coupling operator between the resonator and the simulator. We then apply this scheme to a simulator of the transverse field Ising model implemented with superconducting qubits, where the resonance peaks in the resonator spectrum correspond to the frequencies of the elementary excitations. The effects of resonator damping, qubit decoherence, and resonator backaction are also discussed. This setup can be used to probe a broad range of many-body models.
\end{abstract}
\maketitle 

\section{Introduction\label{sec:intro}}
Quantum simulation, where one controllable quantum system is exploited to emulate another quantum system that is challenging to study by traditional theoretical and experimental techniques, presents us with a powerful platform for studying many-body phenomena~\cite{Feynman, Lloyd}. Over the past two decades, digital and analog quantum simulators for various many-body problems have been proposed. Experimental realization of such simulators in various systems, such as trapped ions, cold atoms or polar molecules, and superconducting quantum circuits, is a fast-growing area~\cite{Cirac+Zoller:12}. Among these systems, superconducting quantum circuits are a promising candidate, given their scalability, long decoherence time, tunable coupling and rich variety of circuit configurations~\cite{squbitReview1, squbitReview2, squbitReview3}. The superconducting circuits include highly-coherent artificial two-level systems (i.e., qubits) and quantum harmonic oscillators (i.e., resonators). These circuit elements can be used to emulate fermionic and bosonic degrees of freedom in a broad range of many-body problems, including quantum spin systems~\cite{Levitov2001, YDWangPRB2007, Garcia-RipollPRB2008, TianPRL2010, GammelmarkNJP2011, MarquardtPRL2013, GellerPRA2013, SolanoPreprint2013, NoriPreprint2013, JQY2011}, coupled cavity array models~\cite{KochPRA2009, PeropadrePRB2013, KSeo2014}, topological phases~\cite{NoriPRA2010, GreentreePRL2012}, and electron-phonon physics~\cite{TianPRB2013,StojanovicPRB2014}. Recent experiments have demonstrated arrays of quantum spins coupled simultaneously to one resonator~\cite{UstinovNCommun2014}, switching of the Chern number on one or two qubits~\cite{LehnertPRL2014, MartinisNature2014} and weak localization of superconducting qubits~\cite{MartinisNCommun2014}.

One important step in implementing quantum simulation is the detection of many-body correlations in the simulators, which gives insights into the many-body systems being emulated. Detection techniques based on specific physical systems have been developed. For cold atoms in an optical lattice, the Bragg scattering technique has been widely used to measure the momentum-space distribution and correlation; for trapped ions, both internal and motional states can be probed with laser pulses. For superconducting systems, circuit quantum electrodynamics (QED), a fruitful approach for manipulating the quantum state of superconducting qubits with a superconducting resonator, provides an advanced readout scheme to detect the quantum states of the qubits~\cite{CQEDtheory, CQEDexp, CQEDdetect1, CQEDdetect2, CQEDdetect3, CQEDdetect4}. Here we study a general framework of a circuit QED scheme, where a superconducting resonator is used as a probe of the many-body properties of a quantum simulator. We show that the spectrum of the resonator is directly related to the correlation function of the coupling operator between the resonator and the simulator, and bears clear fingerprint of the many-body phases of the simulator. Both temporal and spatial correlations of the simulator can be probed with this approach. We illustrate this scheme by applying it to a simulator of the exactly-solvable transverse field Ising model (TFIM). The resonance peaks in the resonator spectrum correspond to the frequencies of the elementary excitations of the TFIM. This scheme can help us understand outstanding many-body problems that cannot be solved by conventional techniques. Depending on the choice of the coupling operator, the spectrum of the resonator can reveal various properties of the simulator, such as spatial correlation in its ground state and the spectrum of its elementary excitations. Furthermore, as superconducting resonators couple to other physical systems, including ensembles of NV centers, trapped ions or electrons and Rydberg atoms, they can also be used to probe quantum simulators implemented with these systems. Our results are extendable to optical cavities which have already been exploited as a probe for many-body effects in cold atoms~\cite{KimbleScience2000, Ritsch2007}. 

This paper is organized as follows. In Sec.~\ref{sec:general}, we study the general framework of the circuit QED scheme on detecting many-body correlations in quantum simulators. In Sec.~\ref{sec:spectrum}, we apply this scheme to a TFIM simulator. The resonator spectrum at finite temperature is derived under various conditions. Effects of resonator damping, qubit decoherence and resonator backaction are discussed in Sec.~\ref{sec:decoherence} and \ref{sec:backaction}. Conclusions are given in Sec.~\ref{sec:conclusions}.

\section{General description\label{sec:general}}
In this section, we study a superconducting resonator coupled to a quantum simulator with a generic Hermitian operator $\hat{Q}$. The Hamiltonian of the resonator mode is $\hat{H}_{c}=\hbar\omega_{c}\hat{a}^{\dagger}\hat{a}+\hat{H}_{\textrm{b}}$, where $\hat{a}$ ($\hat{a}^{\dagger}$) is the annihilation (creation) operator of the mode, $\omega_{c}$ is its frequency, and $\hat{H}_{\textrm{b}}$ represents the interaction between the resonator and its bath modes. We assume that the coupling between the resonator and the simulator has the form 
\begin{equation}
\hat{H}_{\rm{int}}=\hbar\lambda\left(\hat{a}+\hat{a}^{\dagger}\right)\hat{Q}\label{eq:Hint}
\end{equation}
with a coupling strength $\hbar\lambda$. The coupling operator $\hat{Q}$ can have various forms, based on the specifics of the quantum simulator. For example, in \cite{YDWangPRB2007, TianPRL2010}, the coupling operator is a collective operator of all the spins in a spin chain; whereas in \cite{MarquardtPRL2013}, it is an operator of a single qubit or resonator. 

The dynamics of the resonator mode is described by a Heisenberg-Langevin equation~\cite{QuantumOptics}
\begin{equation}
\dot{\hat{a}} = \left(-\textrm{i}\omega_{c}-\kappa/2\right)\hat{a}+\sqrt{\kappa}\hat{a}_{\textrm{in}}(t)-\textrm{i}\lambda\hat{Q}(t),\label{eq:langevin}
\end{equation}
where $\hat{a}_{\textrm{in}}(t)$ is the input noise operator at time $t$ and $\kappa$ is the resonator linewidth (damping rate). For simplicity, we assume the noise correlation as $\langle \hat{a}_{\textrm{in}}(t)\hat{a}_{\textrm{in}}^{\dag}(t^{\prime})\rangle =\left(n_{\textrm{th}}+1\right)\delta(t-t^{\prime})$, which corresponds to a Markovian reservoir with a thermal occupation number $n_{\textrm{th}}$. Meanwhile, we treat the resonator as a weak probe that does not significantly perturb the state of simulator. Under this assumption, the dynamics of the operator $\hat{Q}$ is governed by $\hat{Q}(t)=e^{\textrm{i}\hat{H}_{0}t/\hbar}\hat{Q}e^{-\textrm{i}\hat{H}_{0}t/\hbar}$, where $\hat{H}_{0}$ is the many-body Hamiltonian of the simulator. The backaction of the measurement on $\hat{Q}(t)$ is neglected. We will study the backaction in detail in Sec.~\ref{sec:backaction}. 
 
We solve the above Heisenberg-Langevin equation in the frequency domain. For operator $\hat{o}(t)$, its frequency-domain counterpart is defined as 
\begin{equation}
\hat{o}(\omega)=\int \frac{dt}{\sqrt{2\pi}} e^{\textrm{i}\omega t}\hat{o}(t).\label{eq:oomega}
\end{equation}
Converting Eq.~(\ref{eq:langevin}) to the frequency domain, we derive
\begin{equation}
\hat{a}(\omega)=\frac{-\sqrt{\kappa}\hat{a}_{\textrm{in}}\left(\omega\right)+\textrm{i}\lambda\hat{Q}\left(\omega\right)}{\textrm{i}\left(\omega-\omega_{c}\right)-\kappa/2}\label{eq:aomega}
\end{equation}
together with its conjugate 
\begin{equation}
\hat{a}^{\dag}(\omega)=\frac{-\sqrt{\kappa}\hat{a}_{\textrm{in}}^{\dagger}\left(\omega\right)-\textrm{i}\lambda\hat{Q}\left(\omega\right)}{\textrm{i}\left(\omega+\omega_{c}\right)-\kappa/2},\label{eq:adagomega}
\end{equation}
both of which contain the coupling operator $\hat{Q}\left(\omega\right)$ and the input noise operator $\hat{a}(\omega)$ (or $\hat{a}^{\dag}(\omega)$). The noise operator satisfies $\langle \hat{a}_{\textrm{in}}(\omega)\hat{a}_{\textrm{in}}^{\dag}(\omega^{\prime})\rangle =(n_{\textrm{th}}+1)\delta(\omega+\omega^{\prime})$. 

For the resonator, the time correlation function of its displacement quadrature $\hat{x}=\hat{a}+\hat{a}^{\dag}$ can be written as $C(t)=\left\langle \hat{x}(\tau+t)\hat{x}(\tau)\right\rangle$ for $\tau\gg\left(\kappa\right)^{-1}$. The spectrum of the resonator is then
\begin{equation}
C(\omega)=\int_{-\infty}^{\infty}\frac{dt}{\sqrt{2\pi}}e^{\textrm{i}\omega t-\varepsilon\left|t\right|/2}C(t),\label{eq:Comega}
\end{equation}
where $\varepsilon$ is a small positive number to ensure the convergence of the integral~\cite{MahanBook}. Given Eqs.~(\ref{eq:aomega}-\ref{eq:Comega}), we find that $C(\omega)=C_{\textrm{b}}(\omega) + C_{QQ}(\omega)$. The first term in the spectrum is the contribution of the input noise with
\begin{equation}
\frac{C_{\textrm{b}}(\omega)}{\sqrt{2\pi}}=\left(n_{\textrm{th}}+1\right)f_{\textrm{L}}\left(\omega-\omega_c,\widetilde{\kappa}\right)+n_{\textrm{th}}f_{\textrm{L}}\left(\omega+\omega_c,\widetilde{\kappa}\right),\label{eq:Cbath}
\end{equation}
which contains Lorentzian functions centered at $\omega=\pm\omega_{c}$ with a linewidth $\widetilde{\kappa}=\kappa+\varepsilon$. Here 
\begin{equation}
f_{\textrm{L}}\left(\omega,x\right)=\frac{x}{2\pi\left(\omega^{2}+x^{2}/4\right)},\label{eq:fL}
\end{equation}
and it satisfies $\int_{-\infty}^{\infty} d\omega f_{\textrm{L}}\left(\omega,x\right)=1$. For a resonator mode of, e.g., $12$ GHz frequency, $\hbar\omega_{c}\gg k_{B}T$ at a temperature of $T=20\,\textrm{mK}$. The thermal photon number at these parameters is hence $n_{\textrm{th}}\approx0$, and $C_{\textrm{b}}(\omega)/\sqrt{2\pi}\approx f_{\textrm{L}}(\omega-\omega_c,\widetilde{\kappa})$. The second term in the spectrum is the contribution of the simulator with
\begin{equation}
\frac{C_{QQ}(\omega)}{\sqrt{2\pi}}= \int d\omega_{1}\frac{4\lambda^{2}\omega_{c}^{2}\langle\hat{Q}^{2}(\omega_{1})\rangle f_{\textrm{L}}(\omega-\omega_1,\varepsilon)} {(\kappa^{2}/4+\omega_{c}^{2}-\omega_{1}^{2})^{2}+\kappa^{2}\omega_{1}^{2}},\label{eq:CQQ}
\end{equation}
which is directly associated with the correlation function $\langle\hat{Q}(\omega_1)\hat{Q}(\omega_2)\rangle=2\pi\langle\hat{Q}^{2}(\omega_{1})\rangle\delta(\omega_{1}+\omega_{2})$. By choosing $\omega_{c}$ to be far off resonance from the relevant spectral range of the simulator, $C_{QQ}(\omega)\propto\langle\hat{Q}^{2}(\omega)\rangle$, in the limit of $f_{\textrm{L}}\left(\omega-\omega_1,\varepsilon\right)\rightarrow \delta(\omega-\omega_{1})$. The spectrum of the resonator $C_{QQ}(\omega)$ hence reveals many-body correlations in the quantum simulator. By designing $C_{QQ}(\omega)$ to be much stronger than $C_{\textrm{b}}(\omega)$, the resonator can be a powerful probe of the simulator state. 

The properties of a many-body system are often probed by measuring correlation functions. The scheme studied here can be used to detect various correlation functions, in both temporal and spatial domain, by designing appropriate coupling operator. In Sec.~\ref{sec:spectrum}, we will give an example on detecting the spectrum of the elementary excitations of a many-body system by measuring the resonator spectrum (i.e., temporal correlation). Meanwhile, the equal-time correlation function of the resonator $C(t=0)=\left\langle \hat{x}(\tau+t)\hat{x}(\tau)\right\rangle$ at $t=0$ can be used to study spatial correlations. It can be shown that 
\begin{equation}
C(t=0)\approx\left(2n_{\textrm{th}}+1\right) + \frac{4\lambda^{2}}{\omega_{c}^{2}}\langle\hat{Q}\hat{Q}\rangle,\label{eq:Ct} 
\end{equation}
when $\omega_{c}$ is chosen to be much greater than all other frequency scales in this system. Consider a coupling operator $\hat{Q}(r,r^{\prime})=\hat{A}(r)+\hat{A}(r^{\prime})$, where $\hat{A}(r)$ and $\hat{A}(r^{\prime})$ are operators at the spatial coordinates $r$ and $r^{\prime}$, respectively. Measurement of $C(t=0)$ then yields the spatial correlation function between $\hat{A}(r)$ and $\hat{A}(r^{\prime})$. Such measurement can reveal phase transition in many-body systems, as shown in, e.g., \cite{KSeo2014}.

\section{Application: TFIM simulator\label{sec:spectrum}}
We apply the circuit QED approach to a quantum simulator of the TFIM. The one-dimensional TFIM consists of an array of interacting spin-$1/2$ particles in a transverse external field. The Hamiltonian of this model is 
\begin{equation}
H_{0}=-\hbar J\sum_{i=1}^{N}\hat{\sigma}_{i}^{z}\hat{\sigma}_{i+1}^{z}-\frac{\hbar h_{x}}{2}\sum_{i}^{N}\hat{\sigma}_{i}^{x},\label{eq:H0}
\end{equation}
where $\hbar J$ is the Ising coupling between neighboring spins and $\hbar h_{x}$ is the energy generated by a transverse magnetic field in the $x$-direction. This model (and variations of this model) has been intensively studied as a prototype of continuous quantum phase transition. It is exactly solvable with the Jordan-Wigner transformation, which converts spin particles to fermions. The TFIM can be implemented with various superconducting qubits, such as charge qubit, transmon and flux qubit~\cite{Levitov2001, YDWangPRB2007, TianPRL2010, MarquardtPRL2013}. In Appendix~\ref{aseca}, we give an implementation of this model using superconducting flux qubit. 

A superconducting resonator couples to the TFIM with a coupling operator $\hat{Q}=\sum_{i=1}^{N}\hat{\sigma}_{i}^{x}$. Below we present the spectrum of the resonator under several boundary (or coupling) conditions. 

\subsection{TFIM with periodic boundary condition\label{subsec:periodic}}
Consider a TFIM under periodic boundary condition. We denote the eigenmodes of this model as $\{\hat{\gamma}_{k}\}$ (see Appendix~\ref{asecb} for details). In terms of the eigenmode operators, the coupling operator can be expressed as 
\begin{equation}
\hat{Q}=q_{0}-2\sum_{k}[(u_{k}^{2}-v_{k}^{2})\hat{\gamma}_{k}^{\dagger}\hat{\gamma}_{k}+\textrm{i}u_{k}v_{k}(\hat{\gamma}_{k}^{\dagger}\hat{\gamma}_{-k}^{\dagger}-\hat{\gamma}_{-k}\hat{\gamma}_{k})]\label{eq:Qflux}
\end{equation}
with $q_{0}=N-2\sum_{k} v_{k}^{2}$. For $\hat{Q}(t)=e^{\textrm{i}H_{0}t/\hbar}\hat{Q}e^{-\textrm{i}H_{0}t/\hbar}$, its frequency-domain counterpart contains three components: $\hat{Q}\left(\omega\right)=\hat{Q}_{0}\left(\omega\right)+\hat{Q}_{-}\left(\omega\right)+\hat{Q}_{+}\left(\omega\right)$ with
\begin{subequations}
\begin{align} 
\hat{Q}_{0}\left(\omega\right)&=\sqrt{2\pi}[q_{0}-2\sum_{k}\left(u_{k}^{2}-v_{k}^{2}\right)\hat{\gamma}_{k}^{\dagger}\hat{\gamma}_{k}]\delta\left(\omega\right), \label{eq:Q0}\\
\hat{Q}_{-}\left(\omega\right)&=-2\sqrt{2\pi}\textrm{i}\sum_{k} u_{k}v_{k}\hat{\gamma}_{k}^{\dagger}\hat{\gamma}_{-k}^{\dagger}\delta\left(\omega+2\omega_{k}\right), \label{eq:Q-}\\
\hat{Q}_{+}\left(\omega\right)&=2\sqrt{2\pi}\textrm{i}\sum_{k} u_{k}v_{k}\hat{\gamma}_{-k}\hat{\gamma}_{k}\delta\left(\omega -2\omega_{k}\right) \label{eq:Q+}
\end{align} 
\end{subequations}
at frequencies $\omega=0$ and $\omega=\mp2\omega_{k}$, respectively, where $\omega_{k}$ is the eigenfrequency of mode $\hat{\gamma}_{k}$. Assume that the simulator is in a thermal state with the density matrix $\rho_{s}=e^{-\beta H_{0}}/\textrm{Tr}(e^{-\beta H_{0}})$ ($\beta=1/k_{B}T$) at temperature $T$. The average occupation number of $\hat{\gamma}_{k}$ is $n_{k}=\langle \hat{\gamma}_{k}^{\dag}\hat{\gamma}_{k}\rangle =(e^{\beta\hbar\omega_{k}}-1)^{-1}$. The average of $\langle\hat{\gamma}_{k}^{\dag}\hat{\gamma}_{l}^{\dag}\hat{\gamma}_{p}\hat{\gamma}_{q}\rangle$ can be easily derived using the fermionic commutation relations for the eigenmodes. Using these results, we derive the correlation function $\langle \hat{Q}^{2}(\omega)\rangle$ of the coupling operator.

The simulator contribution $C_{QQ}(\omega)$ can be derived as $C_{QQ}(\omega)=C_{00}(\omega)+C_{nz}(\omega)$. The first term 
\begin{equation}
C_{00}(\omega)=\frac{4\sqrt{2\pi}\lambda^{2}\omega_{c}^{2}Y_{0}f_{\textrm{L}}(\omega,\varepsilon)}{\left(\kappa^{2}/4+\omega_{c}^{2}\right)^{2}}\label{eq:C0}
\end{equation}
is a Lorentzian function centered at $\omega=0$ with
\begin{eqnarray}
Y_{0}&=&q_{0}^{2}+4\sum_{k} \left[\left(u_{k}^{2}-v_{k}^{2}\right)^{2}-q_{0}\left(u_{k}^{2}-v_{k}^{2}\right)\right]n_{k}\nonumber\\
&+&4\sum_{k\ne k^{\prime}}\left(u_{k}^{2}-v_{k}^{2}\right)\left(u_{k^{\prime}}^{2}-v_{k^{\prime}}^{2}\right)n_{k}n_{k^{\prime}}.\label{eq:Y0}
\end{eqnarray}
This term originates from the $\hat{Q}_{0}(\omega)$-component in $\hat{Q}\left(\omega\right)$. The second term in $C_{QQ}(\omega)$ is given by
\begin{equation}
C_{nz}(\omega)=\frac{32\sqrt{2\pi}\lambda^{2}\omega_{c}^{2}f(\omega)}{\left(\kappa^{2}/4+\omega_{c}^{2}- \omega^{2}\right)^{2}+\omega^{2}\kappa^{2}}\label{eq:Cnz}
\end{equation}
with 
\begin{eqnarray}
f(\omega)&=&\sum_{k}u_{k}^{2}v_{k}^{2}n_{k}^{2}f_{\textrm{L}}(\omega+2\omega_{k},\varepsilon) \nonumber\\&+&u_{k}^{2}v_{k}^{2}\left(1-n_{k}\right)^{2}f_{\textrm{L}}(\omega-2\omega_{k},\varepsilon).\label{eq:fomega}
\end{eqnarray}
This term gives resonance peaks centered at $\mp 2\omega_{k}$ and is due to the $\hat{Q}_{\mp}(\omega)$-components in $\hat{Q}\left(\omega\right)$. These peaks correspond to virtual exchange of energy between the resonator mode and a pair of excitations ($\hat{\gamma}_{k}$ and $\hat{\gamma}_{-k}$) in the simulator, and are directly associated with the spectrum of the elementary excitations. The heights of the negative-frequency peaks are proportional to $n_{k}^{2}$, which decreases with the bath temperature and disappears at zero temperature.
 
\begin{figure}
\centering
\includegraphics[clip,width=\columnwidth]{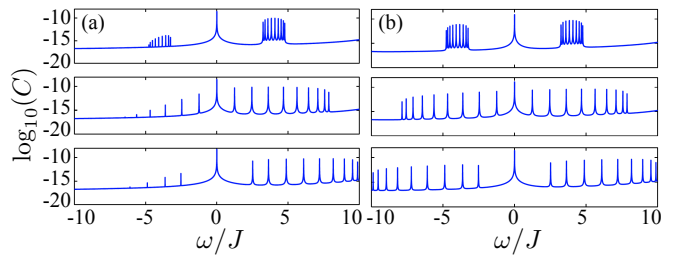}
\caption{The logarithmic spectrum $\log_{10}(C)$ verses $\omega/J$ at (a) $T=20$ mK and (b) $T=100$ mK. The panels from top to bottom are for $h_{x}/2J=0.2,\,1,\,1.5$, respectively.}
\label{fig1}
\end{figure}
\begin{figure}
\centering
\includegraphics[clip,width=\columnwidth]{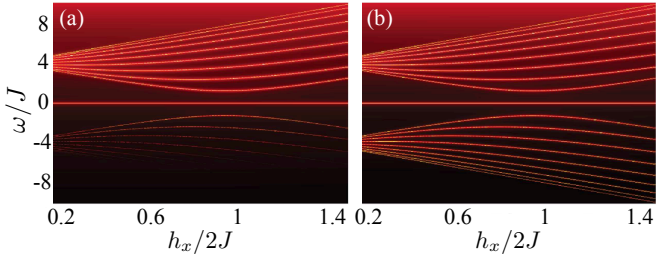}
\caption{The logarithmic spectrum $\log_{10}(C)$ versus $h_{x}/2J$ and $\omega/J$ at (a) $T=20$ mK and (b) $T=100$ mK.}
\label{fig2}
\end{figure}
We choose the following parameters for the superconducting resonator and the TFIM simulator: $\omega_{c}/2\pi=12$ GHz, $\kappa/2\pi=100$ kHz, $\varepsilon/2\pi=600$ kHz, $J/2\pi=1$ GHz and $\lambda/2\pi=40$ MHz. The transverse filed $h_{x}/2\pi$ varies between $400$ MHz and $3$ GHz, i.e., $h_{x}/2J$ varies between $0.2$ and $1.5$, including the quantum critical point $h_{x}^{c}/2J=1$. The total spectrum of the resonator is plotted in Figs.~\ref{fig1}(a) and \ref{fig1}(b) using these parameters at selected values of $h_{x}/2J$ and temperature. For a qubit array of size $N=20$, there are a total of nine peaks at positive (negative) frequency. The peak positions are at $\omega=2\omega_{k}$ ($\omega=-2\omega_{k}$) with $k=\pm 2\pi n /N$ and $n=1-9$. For $k=0$ and $k=\pi$, $C_{nz}(\omega_{k})=0$ with $u_{k}v_{k}=0$, and the total spectrum $C(\omega_{k}=0)$ is due to the contribution of $C_{00}$ and $C_{\textrm{b}}$. Some negative-frequency peaks are illegible due to their small magnitude, as their heights decrease quickly when the temperature decreases. The sharp peak at $\omega=0$ comes mainly from $C_{00}$. At $h_{x}\ll 2J$, the frequencies of the elementary excitations are centered around $2J$, and the peaks are located densely around $\mp4J$. As $h_{x}$ increases, the eigenfrequencies, and hence the peaks, span over a wider range. Our numerical result is well explained by the solution given by Eqs.~(\ref{eq:C0}) and (\ref{eq:Cnz}). Note that with these parameters, the contribution of the bath modes $C_{\textrm{b}}(\omega)$ is dominated by a sharp peak at $\omega=\omega_{c}$, which is far off resonance from the peaks at $\pm 2\omega_{k}$. The $C_{\textrm{b}}(\omega)$ term hence only generates a smooth-varying spectral background within the range of these peaks, which is a few order of magnitude smaller than $C_{QQ}$. In Fig.~\ref{fig2}, we plot $\log_{10}C(\omega)$ for $h_{x}\in[0.2,1.5)$ at selected temperatures. It can be seen that the resonance peaks exactly reflect the excitation spectrum of the TFIM with the excitation gap approaching zero at the critical point $h_{x}^{c}/2J=1$. 
 
\begin{figure}
\centering
\includegraphics[clip,width=\columnwidth]{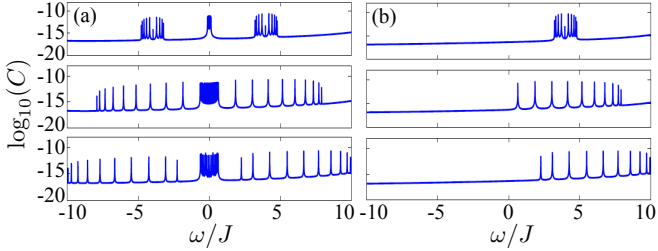}
\caption{The logarithmic spectrum $\log_{10}(C)$ verses $\omega/J$ at (a) $T=100$ mK and (b) $T=0$ mK with site-dependent coupling. The panels from top to bottom are for $h_{x}/2J=0.2,\,1,\,1.5$, respectively.}
\label{fig3}
\end{figure}
\subsection{TFIM with site-dependent coupling\label{subsec:sitedependence}}
In Sec.~\ref{subsec:periodic}, the qubits in the TFIM span a very small segment of the resonator, where the resonator field (and the coupling strength) can be assumed uniform. For a TFIM spanning a large range of the resonator, the coupling becomes site-dependent. Let neighboring qubits be equally spaced over the entire length $L$ of a resonator. For the lowest even mode of the resonator, its magnetic field is $B(x)=B_{0}\sin(2\pi x/L)(\hat{a}+\hat{a}^{\dag})$. The coupling operator is then $\hat{Q}=\sum_{i=1}^{N}\sin(2\pi i/N)\hat{\sigma}_{i}^{x}$. In the frequency domain, the coupling operator is then
\begin{eqnarray}
\hat{Q}(\omega)&=&\textrm{i}\sqrt{2\pi}\sum_{k} [ \label{eq:Qspatialomega} \\ 
&+&(u_ku_{\bar{k}}\hat{\gamma}_k^{\dag}\hat{\gamma}_{\bar{k}}+v_{k}v_{\bar{k}}\hat{\gamma}_{-k}^{\dag}\hat{\gamma}_{-\bar{k}})\delta(\omega+\omega_k-\omega_{\bar{k}})\nonumber\\ 
&-&(u_ku_{\bar{k}}\hat{\gamma}_{\bar{k}}^{\dag}\hat{\gamma}_{k}+v_kv_{\bar{k}}\hat{\gamma}_{-\bar{k}}^{\dag}\hat{\gamma}_{-k})\delta(\omega-\omega_k+\omega_{\bar{k}})\nonumber\\ 
&+&\textrm{i}(u_kv_{\bar{k}}\hat{\gamma}_k^{\dag}\hat{\gamma}_{-\bar{k}}^{\dag}+v_ku_{\bar{k}}\hat{\gamma}_{-k}^{\dag}\hat{\gamma}_{\bar{k}}^{\dag})\delta(\omega+\omega_k+\omega_{\bar{k}})\nonumber\\
&-&\textrm{i}(v_ku_{\bar{k}}\hat{\gamma}_{-k}\hat{\gamma}_{\bar{k}}+u_kv_{\bar{k}}\hat{\gamma}_{k}\hat{\gamma}_{-\bar{k}})\delta(\omega-\omega_k-\omega_{\bar{k}})]\nonumber 
\end{eqnarray}
with $\bar{k}=k-2\pi/N$ ($\bar{k}=\pi$ for $k=-\pi+2\pi/N$). The first (second) term in $\hat{Q}(\omega)$ corresponds to the absorption (emission) of a particle with momentum $\pm \bar{k}$ and the simultaneous emission (absorption) of another particle with momentum $\pm k$. The third (fourth) term corresponds to the emission (absorption) of a pair of particles at $\pm k$ and $\mp\bar{k}$. Hence $\hat{Q}(\omega)$ contains frequency components of $\omega=\mp (\omega_{k}-\omega_{\bar{k}})$ and $\omega=\mp (\omega_{k}+\omega_{\bar{k}})$. 

The simulator contribution $C_{QQ}$ in the resonator spectrum can be derived as
\begin{equation}
C_{QQ}=\frac{4\sqrt{2\pi}\lambda^2\omega_c^2 f_{\textrm{sd}}(\omega)}{(\kappa^2/4+\omega_c^2-\omega^2)^2+\omega^2\kappa^2}\label{eq:CQQspatial}
\end{equation}
with 
\begin{eqnarray}
f_{\textrm{sd}}(\omega)&=&\sum[M_kn_k(1-n_{\bar{k}})f_{\textrm{L}}(\omega+\omega_k-\omega_{\bar{k}},\varepsilon)\nonumber\\&+&M_k(1-n_k)n_{\bar{k}}f_{\textrm{L}}(\omega-\omega_k+\omega_{\bar{k}},\varepsilon)\nonumber\\
&+&P_kn_kn_{\bar{k}}f_{\textrm{L}}(\omega+\omega_k+\omega_{\bar{k}},\varepsilon)\label{eq:fsd}\\ 
&+&P_k(1-n_k)(1-n_{\bar{k}})f_{\textrm{L}}(\omega-\omega_k-\omega_{\bar{k}},\varepsilon)],\nonumber
\end{eqnarray}
$M_k=u_k^2u_{\bar{k}}^2+v_k^2v_{\bar{k}}^2$ and $P_k=(u_kv_{\bar{k}}+v_{k}u_{\bar{k}})^2$. In Figs.~\ref{fig3}(a) and \ref{fig3}(b), the spectrum of the resonator is plotted. The peaks at $\omega=\mp (\omega_{k}+\omega_{\bar{k}})$ correspond to the absorption or emission of a pair of excitations at $\pm k$ and $\mp\bar{k}$. The heights of the negative-frequency peaks are proportional to $n_{k}n_{\bar{k}}$, which disappear at zero temperature. In Fig.~\ref{fig3}(a), one observes densely-packed resonance peaks near $\omega=0$ with frequencies $\omega=\mp (\omega_{k}-\omega_{\bar{k}})$, which correspond to the emission (absorption) of one excitation at $\pm k$ and the simultaneous absorption (emission) of another excitation at $\pm\bar{k}$. The heights of these peaks are proportional to $n_{k}$ or $n_{\bar{k}}$, which disappear at zero temperature, as shown in Fig.~\ref{fig3}(b). The lattice-dependent coupling hence enables the exchange between neighboring momentum-space modes. 

\subsection{TFIM with open boundary condition\label{subsec:openboundary}}
In experiments, it is often more practical to realize a simulator with small lattice size and open boundary condition. Here we calculate the spectrum of the resonator with the TFIM under open boundary condition. The Hamiltonian of a TFIM under open boundary condition has the bilinear form 
\begin{equation}
\hat{H}_{0}=\hbar\sum_{i,j=1}^{N}[\hat{c}_{i}^{\dagger}A_{ij}\hat{c}_{j}+1/2(\hat{c}_{i}^{\dagger}B_{ij}\hat{c}_{j}^{\dagger}+h.c.)]\label{eq:quadH}
\end{equation}
with the matrices
\begin{equation}
A=\left(\begin{array}{cccc}
h_x& -J & \cdots & 0\\
-J & h_x &  & 0\\
\vdots &  & \ddots & -J\\
0 & 0 & -J & h_x
\end{array}\right);\quad
B=\left(\begin{array}{cccc}
0 & -J & \cdots & 0\\
J & 0 &  & 0\\
\vdots &  & \ddots & -J\\
0 & 0 & J & 0
\end{array}\right),\label{eq:pbcB}
\end{equation}
where the qubits at sites $1$ and $N$ only couple to one neighboring qubit. Following the approach in \cite{LiebAnnPhys1961}, this Hamiltonian can be diagonalized as $\hat{H}_{0}=\sum \hbar\omega_{m}\hat{\eta}_{m}^\dagger \hat{\eta}_{m}$ with eigenmodes $\hat{\eta}_{m}=\sum_{i=1}^{N}(g_{mi}\hat{c}_{i}+h_{mi}\hat{c}_{i}^{\dagger})$, where $g_{mi},\,h_{mi}$ are coefficients for mode $\hat{\eta}_{m}$. In the frequency domain, the coupling operator is then 
\begin{eqnarray}
\hat{Q}(\omega)&=&\sqrt{2\pi}N\delta(\omega)\nonumber\\
&-&2\sqrt{2\pi}\sum[g_{mi}g_{ni}^{\star}\hat{\eta}_{m}^\dag\hat{\eta}_{n}\delta(\omega+\omega_m-\omega_n)\nonumber\\
&+&h_{mi}h_{ni}^{\star}\hat{\eta}_{m}\hat{\eta}_{n}^\dag\delta(\omega-\omega_m+\omega_n)\nonumber\\
&+&g_{mi}h_{ni}^{\star}\hat{\eta}_{m}^\dag\hat{\eta}_{n}^{\dag}\delta(\omega+\omega_m+\omega_n)\label{eq:QomegaOpen}\\
&+&h_{mi}g_{ni}^{\star}\hat{\eta}_{m}\hat{\eta}_{n}\delta(\omega-\omega_m-\omega_n)],\nonumber
\end{eqnarray}
which includes frequency components of $\omega=0$ and $\omega=\pm(\omega_{m}\pm\omega_{n})$.

\begin{figure}
\centering
\includegraphics[clip,width=\columnwidth]{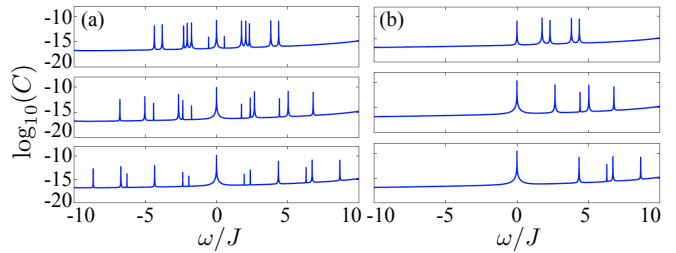}
\caption{The logarithmic spectrum $\log_{10}(C)$ verses $\omega/J$ at (a) $T=100$ mK and (b) $T=0$ mK for a small array of size $N=4$ under open boundary condition. The panels from top to bottom are for $h_{x}/2J=0.2,\,1,\,1.5$, respectively.}
\label{fig4}
\end{figure}

The simulator contribution in the resonator spectrum can be decomposed as $C_{QQ}(\omega)=C_{00}+C_{nz}$. Here
\begin{equation}
C_{00}=\frac{4\sqrt{2\pi}\lambda^2\omega_c^2T_{00}f_{\textrm{L}}(\omega,\varepsilon)}{(\kappa^2/4+\omega_c^2)^2}, \label{eq:C00Open}
\end{equation}
centering at $\omega=0$ with $T_{00}$ being a constant that depends on $g_{mi}$ and $h_{mi}$; and $C_{nz}$ has the form
\begin{equation}
C_{nz}=\frac{16\sqrt{2\pi}\lambda^2\omega_c^2f_{\textrm{op}}(\omega)}{(\kappa^2/4+\omega_c^2-\omega^2)^2+\omega^2\kappa^2}\label{eq:CnzOpen}
\end{equation}
with 
\begin{eqnarray}
f_{\textrm{op}}(\omega)&=&\sum [T_{mn}^{+-}f_{\textrm{L}}(\omega+\omega_m-\omega_n,\varepsilon) \nonumber \\
&+&T_{mn}^{++}f_{\textrm{L}}(\omega+\omega_m+\omega_n,\varepsilon)\nonumber\\
&+& T_{mn}^{--}f_{\textrm{L}}(\omega-\omega_m-\omega_n,\varepsilon)],\label{eq:fLopen}
\end{eqnarray}
where $T_{mn}^{+-}$, $T_{mn}^{++}$ and $T_{mn}^{--}$ depend on  $g_{mi}$, $h_{mi}$, and the thermal occupation numbers $n_{m,n}$. The resonator spectrum for a small array of $N=4$ is plotted in Fig.~\ref{fig4}. The resonance peaks correspond to the frequencies $\pm(\omega_{m}\pm\omega_{n})$ for the eigenmodes $\hat{\eta}_{m,n}$. For $N=4$ and $h_{x}/2J=0.2$, the eigenvalues are $\left\{ \omega_{1-4}\right\} =\{0.003,1.754, 2.059,2.308\}J$. At zero temperature, we see four resonance peaks at positive frequency, corresponding to $\{\omega_1+\omega_2, \omega_2+\omega_3, \omega_3+\omega_4, \omega_4+\omega_1\}$, respectively. The magnitude of these peaks is determined by the matrix element of the eigenmodes and $n_{m,n}$. 

\section{Decoherence\label{sec:decoherence}}
The superconducting resonator and the qubits in the TFIM simulator are subject to the disturbance of environmental modes. For the resonator, we use an input noise operator to model its coupling with bath modes. The input noise contributes a term $C_{\textrm{b}}(\omega)$, which is a Lorentzian function centered at the frequency of the resonator $\omega_{c}$, to the total spectrum. By choosing the resonator frequency to be a few gigahertz away from the peak positions in $C_{QQ}(\omega)$, the contribution of the resonator noise is a few orders of magnitude smaller than the contribution of the simulator. The decoherence of the qubits is considered implicitly in our discussion. We assume that the simulator is in contact with a thermal bath at temperature $T$, and is subject to thermal fluctuations of the bath modes. The spectrum of the resonator is calculated at finite temperature. This approach treats the decoherence of the qubits phenomenologically, which is sufficient for the discussion in this work.

\section{Resonator backaction\label{sec:backaction}}
In the previous sections, we assume that the time dependence of $\hat{Q}(t)$ is governed entirely by the simulator Hamiltonian, and we neglect the measurement backaction on the quantum simulator. Below we estimate the backaction on the TFIM simulator with a perturbative approach. Our method can be extended to a general study of measurement backaction on other simulators. 

We treat the coupling between the simulator and the resonator as a perturbation on the unperturbed Hamiltonian $\hat{H}_{s}=\hbar\omega_{c}\hat{a}^{\dag}\hat{a}+\hat{H}_{0}$. The coupling Hamiltonian $\hat{H}_{\rm{int}}=\hbar\lambda(\hat{a}+\hat{a}^{\dag})\hat{Q}$ induces nonzero transition matrix elements between eigenstates of the unperturbed system. All these transitions involve emission or absorption of one microwave photon. To ensure the validity of the perturbative approach, the transition matrix elements need to be much smaller than the energy separation between the initial and the final states of the transition. In our system,  the energy separation is always greater than $[\omega_{c}-(4J+2h_{x})]$ when $\omega_{c}\gg J, h_{x}$; whereas the transition matrix elements are upper bounded by $\lambda \sqrt{\langle\hat{a}\hat{a}^{\dag}\rangle} |\hat{Q}|$ with $|\hat{Q}|\lesssim N/2$. Using Eq.~(\ref{eq:langevin}), we have
\begin{equation}
\langle\hat{a}\hat{a}^{\dag}\rangle=1+(i\lambda/\kappa)\langle\sum_{i}\hat{\sigma}_{i}^{x}(\hat{a}-\hat{a}^{\dag})\rangle\le1+(\lambda N/\omega_{c})^{2}.\label{nave}
\end{equation}
It can then be shown that this perturbative approach will be self-consistent when $\lambda N/2<[\omega_{c}-(4J+2h_{x})]$, which puts an upper bound on the size of simulator. With the parameters in Sec.~\ref{subsec:periodic} and $h_{x}=J$, it gives $N\lesssim300$. 

Now we derive the second-order perturbative correction to the simulator Hamiltonian. Note that because all diagonal matrix elements of $\hat{H}_{\rm{int}}$ are zero, there is no first order correction to the Hamiltonian. The dominant second-order correction can be written as 
\begin{eqnarray}
\delta H_{I}^{(2)}&\approx&-\frac{4\hbar\lambda^{2}q_{0}}{\omega_{c}}\sum_{k}\cos(2\theta_{k})\hat{\gamma}_{k}^{\dagger}\hat{\gamma}_{k}\nonumber\\
&+&\frac{4\hbar\lambda^{2}}{\omega_{c}}\left(\sum_{k}\cos(2\theta_{k})\hat{\gamma}_{k}^{\dagger}\hat{\gamma}_{k}\right)^{2},\label{2ndorder}
\end{eqnarray}
where $\theta_{k}$ is defined in Eq.~(\ref{eq:theta}), and $q_{0}=\sum_{k}\cos(2\theta_{k})$ increases from zero to $N$ as $h_{x}$ increases. For simplicity, we set $q_{0}\sim N/2$. The backaction of the measurement hence generates a shift $\delta \omega_{k}$ on the frequency of the eigenmodes with $\vert\delta \omega_{k}\vert\sim 2\lambda^{2}N/\omega_{c}$.

For the backaction to be negligible, $\delta \omega_{k}$ needs to be smaller than the frequency spacing $\Delta\omega_{k}=|\partial \omega_{k}/\partial k|\Delta k$ ($\Delta k=2\pi/N$ for periodic boundary condition) between adjacent eigenmodes. At small transverse field $h_{x}\ll 2J$, Eq.~(\ref{eq:omK}) gives $\omega_{k}\approx 2J-h_{x}\cos(k)$ with a minimal frequency spacing of $\Delta\omega_{k}=h_{x}\Delta k^{2}/2$ for $k=0$. To satisfy $\vert\delta \omega_{k}\vert<\Delta\omega_{k}$, it requires $\lambda^{2}N/\omega_{c}< \pi^{2}h_{x}/N^{2}$; at $h_{x}/2J=0.5$, for example, this relation becomes $N< 40$ with our parameters. At $h_{x}=2J$ (the critical point), $\omega_{k}=4J|\sin(k/2)|$, and the frequency spacing is $\Delta\omega_{k}=4\pi J/N$ for $k=0$. It then requires $\lambda^{2}N/\omega_{c}< 2\pi J/N$, which gives $N< 217$ with our parameters. At strong transverse field $h_{x}\gg 2J$, $\omega_{k}\approx h_{x}-2J\cos(k)$; the condition becomes $\lambda^{2}N/\omega_{c}< 2\pi^{2} J /N^{2}$. This gives $N< 150$ with our parameters. Our analysis hence shows that the resonator backaction on the TFIM simulator can be neglected in a moderate-size array.

\section{Conclusions\label{sec:conclusions}}
In summary, we study a circuit QED setup for probing the many-body properties of a quantum simulator with a superconducting resonator. We show that the spectrum of the resonator is dominated by the correlation function of the coupling operator between the resonator and the simulator. By designing appropriate resonator-simulator coupling, various properties of the simulator can be revealed by measuring the resonator spectrum. We illustrate this scheme by calculating the spectrum of a resonator coupled to an exactly solvable model, a TFIM simulator, where the resonator spectrum can be mapped to that of the elementary excitations of the TFIM. 

This approach can be used to probe many-body correlations in a broad range of quantum simulators and provide answers to problems that cannot be solved by conventional theoretical or experimental techniques. Both temporal and spatial correlation functions of a simulator can be obtained from the resonator. Such scheme thus provides insights into outstanding questions in many-body physics. For example, by coupling a resonator to a Fermi-Hubbard model via properly designed operators, the spin and charge orders of this model can be probed, which could help us understand quantum magnetism and high-Tc superconductivity. With conventional techniques, this model can only be exactly solved in one dimension and infinite dimension. The resonator can also be coupled to a multiconnected Jaynes-Cummings lattice model, which could be a promising system for observing quantum phase transition in cavity polaritons~\cite{KSeo2014}. With a coupling operator $\hat{Q}(i,j)=\sigma_{i}^{x}+\sigma_{j}^{x}$ for the qubits on sites $i$ and $j$, the equal-time correlation function $C(t=0)$ tells us the spatial correlation in this model and predicts the critical points in the transition between the Mott insulator and the superfluid phases.

\section*{Acknowledgements\label{sec:Acknowledgement}}
L.H.D. and L.T. are supported by the National Science Foundation under Award Numbers 0956064 and 0916303. J.Q.Y. is supported by the NSFC Grant No. 91421102, the MOST 973 Program Grant No. 2014CB921401, and the NSAF Grant No. U1330201. L.T. thanks Fudan University for hospitality during her visit.

\appendix
\section{TFIM simulator with flux qubits\label{aseca}}
\begin{figure}
\centering
\includegraphics[clip,width=0.9\columnwidth]{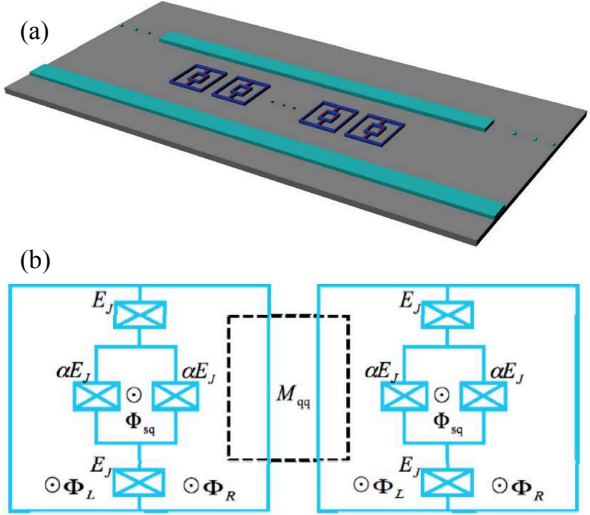}
\caption{(a) Schematic of a TFIM simulator made of an array of superconducting flux qubits coupled to a coplanar waveguide resonator. (b) The coupling between neighboring qubits via mutual inductance.}
\label{fig5} 
\end{figure}

Here we show that superconducting flux qubits can be used to form a TFIM, cf. Fig.~\ref{fig5}(a), and generate the qubit-resonator coupling used in Sec.~\ref{sec:spectrum}. The superconducting flux qubit, also known as the persistent-current qubit, is typically made of four Josephson junctions connected in superconducting loops~\cite{fluxqubit, squbitReview4}. As shown in Fig.~\ref{fig5}(b), two junctions in the circuit have Josephson energy $E_{J}$, while the other two junctions have Josephson energy $\alpha E_{J}$ and form a dc SQUID. We denote the external magnetic flux in the left (right) main loop as $\Phi_{L}$ ($\Phi_{R}$), the external flux in the SQUID loop as $\Phi_{\textrm{sq}}$, and the gauge-invariant phase difference of the top (bottom) junction as $\varphi_{t}$ ($\varphi_{b}$). The total Josephson energy is then
\begin{eqnarray}
U_{J}&=&-E_{J}[\cos\left(\varphi_{t}\right)+\cos\left(\varphi_{b}\right)]\nonumber\\
&-&2\alpha E_{J}\cos\left(\pi f_{\textrm{sq}}\right)\cos\left(\varphi_{t}+\varphi_{b}+\pi f_{\textrm{d}}\right),\label{eq:UJ}
\end{eqnarray}
where $f_{\textrm{sq}}=\Phi_{\textrm{sq}}/\Phi_{0}$, $f_{\textrm{d}}=\left(\Phi_{R}-\Phi_{L}\right)/\Phi_{0}$, and $\Phi_{0}=h/2e$ is the flux quantum. The qubit states are persistent-current states with opposite circulating currents $\pm I_{\textrm{cir}}$, and the energy splitting between these states can be controlled by the flux difference $f_{\textrm{d}}$ in the main loops. Let the persistent-current states be eigenstates of the operator $\hat{\sigma}_{i}^{z}$ of the $i$-th qubit. We assume that the qubits are biased at the degeneracy point with $f_{\textrm{d}}=1$, where the energies of the persistent-current states are equal to each other. The bare qubit Hamiltonian summed over all sites is then $\hat{H}_{\textrm{tf}}=(\hbar h_{x}/2)\sum \hat{\sigma}_{i}^{x}$, where $\hbar h_{x}$ is the quantum tunneling between the persistent-current states and can be controlled by varying the flux $f_{\textrm{sq}}$ in the SQUID loop. Neighboring qubits couple via the mutual inductance between their main loops as indicated in Fig.~\ref{fig5}(b). The inductive coupling between the SQUID loop of one qubit and the main loops of its neighboring qubits is designed to be negligible. The qubit-qubit coupling gives an interaction $\hat{H}_{\textrm{qq}}=-\hbar J\sum_{i=1}^{N}\hat{\sigma}_{i}^{z}\hat{\sigma}_{i+1}^{z}$, where $\hbar J=M_{\textrm{qq}}I_{\textrm{cir}}^{2}$ is the inductive energy in terms of the mutual inductance $M_{\textrm{qq}}$ and the circulating current $I_{\textrm{cir}}$. The Hamiltonian for the qubit array is then $\hat{H}_{0}=\hat{H}_{\textrm{tf}}+\hat{H}_{\textrm{qq}}$, which has the form of the TFIM in Eq.~(\ref{eq:H0}). As demonstrated in recent experiments, $h_{x}/2\pi$ can be adjusted over a broad range of frequencies between sub-gigahertz to a few gigahertz~\cite{MITexp2011}, and $J/2\pi $ can easily reach gigahertz. 

Coupling between superconducting flux qubits and superconducting resonator has been demonstrated experimentally~\cite{CQEDfluxqubit1, CQEDfluxqubit2, CQEDfluxqubit3, CQEDfluxqubit4, CQEDfluxqubit5}. Here consider a coplanar waveguide resonator inductively coupled to the flux qubits in the TFIM. The magnetic field of the resonator threads through the main loops and the SQUID loop of each qubit and modifies the total external flux in the loops. Let the area of the main loops be equal to each other. The resonator field generates the same amount of flux in these loops with $\Phi_{L}\rightarrow\Phi_{L}+\delta\Phi_{\textrm{m}} \left(\hat{a}+\hat{a}^{\dag}\right)$ and $\Phi_{R}\rightarrow\Phi_{R}+\delta\Phi_{\textrm{m}} \left(\hat{a}+\hat{a}^{\dag}\right)$. The flux difference $f_{\textrm{d}}$ is hence not affected by the presence of the resonator. The resonator field modifies the flux in the SQUID loop as $\Phi_{\textrm{sq}}\rightarrow \Phi_{\textrm{sq}} + \delta\Phi_{\textrm{sq}}\left(\hat{a}+\hat{a}^{\dag}\right)$, which generates a qubit-resonator coupling. Using Eq.~(\ref{eq:UJ}), we derive the coupling as 
\begin{equation}
\hat{H}_{int}=-s_{2}E_{J}\left(\delta\Phi_{\textrm{sq}}/\Phi_{0}\right)\left(\hat{a}+\hat{a}^{\dagger}\right)\sum_{i=1}^{N}\hat{\sigma}_{i}^{x},\label{eq:Hintflux}
\end{equation}
which gives a coupling operator $\hat{Q}=\sum_{i=1}^{N}\hat{\sigma}_{i}^{x}$ with coupling strength $\hbar\lambda = -s_{2}E_{J}(\delta\Phi_{\textrm{sq}}/\Phi_{0})$. Here $s_{2}$ is a numerical coefficient. At $\alpha=0.8$, $s_{2}\sim0.2$ for typical flux qubits. With $\delta\Phi_{\textrm{sq}}=10^{-3}\,\Phi_{0}$ and $E_{J}/2\pi\hbar=200\,\textrm{GHz}$, we have $\lambda/2\pi \sim 40\,\textrm{MHz}$. The coupling strength is far below the energy scales of the TFIM, in which both $h_{x}$ and $J$ are in the gigahertz range.

\section{TFIM under periodic boundary condition\label{asecb}}
The TFIM can be solved by the Jordan-Wigner transformation (JWT)~\cite{Sachdev}: 
\begin{equation}
\hat{\sigma}_{i}^{z}=(\hat{c}_{i}^{\dagger}+\hat{c}_{i})\prod_{j=1}^{i-1}(1-2\hat{c}_{j}^{\dagger}\hat{c}_{j}),\quad\hat{\sigma}_{i}^{x}=(1-2\hat{c}_{i}^{\dagger}\hat{c}_{i}), \label{eq:JWT}
\end{equation}
where $\hat{c}_{i}$ is the annihilation operator of a spinless fermion at site $i$. The Hamiltonian then becomes
\begin{equation}
\hat{H}_{0}=-\hbar J\sum_{i=1}^{N}(\hat{c}_{i}^{\dagger}\hat{c}_{i+1}^{\dagger}+\hat{c}_{i}^{\dagger}\hat{c}_{i+1}+h.c.)+\hbar h_{x}\sum_{i}^{N}\hat{c}_{i}^{\dagger}\hat{c}_{i}\label{eq:cfermion}
\end{equation}
with $\hat{c}_{N+1}=\hat{c}_{1}$. Here the extra term at the boundary of the spin chain is neglected in the limit of large $N$. This Hamiltonian is bilinear and can be exactly diagonalized. In the momentum space, 
\begin{equation}
\hat{H}_{0}=\sum_{k}\hbar\left(h_{x}-2J\cos k\right)\hat{c}_{k}^{\dagger}\hat{c}_{k}-\hbar J\sum_{k}(\hat{c}_{k}^{\dagger}\hat{c}_{-k}^{\dagger}e^{\textrm{i}k}+h.c.),\label{eq:cfermion-k}
\end{equation}
where $\hat{c}_{k}=\sum_{i}e^{-\textrm{i}ki}\hat{c}_{i}/\sqrt{N}$ for $k=2\pi m_{k}/N$ ($-N/2<m_{k}\le N/2$). Using the Bogoliubov transformation
\begin{equation}
\hat{c}_{k}=u_{k}\hat{\gamma}_{k}+iv_{k}\hat{\gamma}_{-k}^{\dag},\label{eq:bt}
\end{equation}
where $\hat{\gamma}_{k}$'s are fermionic operators with the coefficients $u_{k}=\cos\theta_{k}$, $v_{k}=\sin\theta_{k}$ and 
\begin{equation}
\tan(2\theta_{k})=\frac{2J\sin k}{h_{x}-2J\cos k},\label{eq:theta}
\end{equation}
the TFIM Hamiltonian becomes $\hat{H}_{0}=\sum_{k}\hbar\omega_{k}\hat{\gamma}_{k}^{\dag}\hat{\gamma}_{k}+E_{g}$. Here $E_{g}$ is the ground state energy. The frequencies of the elementary excitations are
\begin{equation}
\hbar\omega_{k}=2\hbar J\sqrt{1+\left(h_{x}/2J\right)^{2}-\left(h_{x}/J\right)\cos k}.\label{eq:omK}
\end{equation}

\end{document}